# Gender differences of the effect of vaccination on perceptions of COVID-19 and mental health in Japan

Eiji Yamamura[1*], Youki Kosaka[2], Yoshiro Tsutsui[3], Fumio Ohtake[4],

[1] Department of Economics, Seinan Gakuin University, Fukuoka, Japan [2] Kyoto Economic College, Japan [3] Kyoto-Bunkyo University, Japan [4] Osaka University, Japan

## Abstract

Vaccination has been promoted to mitigate the spread of the coronavirus disease 2019 (COVID-19). Vaccination is expected to reduce the probability of and alleviate the seriousness of COVID-19 infection. Accordingly, this might significantly change an individual's subjective well-being and mental health. However, it is unknown how vaccinated people perceive the effectiveness of COVID-19 and how their subjective well-being and mental health change after vaccination. We thus observed the same individuals on a monthly basis from March 2020 to September 2021 in all parts of Japan. Then, large sample panel data (N=54,007) were independently constructed. Using the data, we compared the individuals' perceptions of COVID-19, subjective well-being, and mental health before and after vaccination. Furthermore, we compared the effect of vaccination on the perceptions of COVID-19 and mental health for females and males. We used the fixed-effects model to control for individual time-invariant characteristics. The major findings were as follows: First, the vaccinated people perceived the probability of getting infected and the seriousness of COVID-19 to be lower than before vaccination. This was observed not only when we used the whole sample, but also when we used sub-samples. Second, using the whole sample, subjective well-being and mental health improved. The same results were also observed using the sub-sample of females, whereas the improvements were not observed using a sub-sample of males.

# 1. Introduction

Vaccination against the coronavirus disease 2019 (COVID-19) is anticipated to play a critical role in mitigating the spread of COVID-19. Many newly reported cases of COVID-19 have been reduced in countries where vaccines have become rapidly pervasive(WHO Coronavirus (COVID-19) Dashboard 2021). Through scientific experiments, the COVID-19 vaccine reduced the probability of infection and the seriousness of COVID-19. The sufficient rate of the vaccinated population in society must reach herd immunity to terminate the COVID-19 pandemic(Randolph and Barreiro 2020). However, some individuals hesitate to receive the COVID-19 vaccine(Almaghaslah et al. 2021a; Lucia, Kelekar, and Afonso 2021a; Machingaidze and Wiysonge 2021a; Murphy et al. 2021a; Solís Arce et al. 2021). Their attitude may change if they know that vaccinated people have a more positive view about the vaccination after receiving the vaccine. Therefore, how and the extent to which the subjective views about the effectiveness of the COVID-9 vaccine changes after one gets vaccinated should be examined.

Various measures against COVID-19, such as lockdown restrictions, cause significant economic loss(Inoue, Murase, and Todo 2021; Mottaleb, Mainuddin, and Sonobe 2020) and exert a detrimental impact on individuals' mental health(Chinna et al.

2021; Fiorenzato et al. 2021; Greyling, Rossouw, and Adhikari 2021; Ogden 2021). In Japan, even without enforcement, individuals voluntarily exhibit preventive behaviors, such as staying indoors and avoiding face-to-face conversations(Muto et al. 2020; Watanabe and Yabu 2021; Yamamura and Tsutsui 2020). Accordingly, this changed lifestyle, for instance, lack of exercise and short sleep duration, results in a decline in mental health(Nagasu and Yamamoto 2020; Yamamura and Tsustsui 2021a; Yamamura and Tsutsui 2020). Vaccination is anticipated to reduce the probability of contracting COVID-19; thus, vaccinated individuals can return to normal daily life. This return to normal daily life improves subjective well-being and mental health, so vaccination for people with mental illness is necessary(Mazereel et al. 2021a, 2021b; Siva 2021; Warren, Kisely, and Siskind 2021).

The mental conditions of vaccinated individuals improved in the U.S(Perez-Arce et al. 2021). However, hesitancy to be vaccinated was observed in various countries(Almaghaslah et al. 2021b; Machingaidze and Wiysonge 2021b; Murphy et al. 2021b; Solís Arce et al. 2021). This has hampered the establishment of herd immunity and increased social costs caused by COVID-19. Furthermore, people are less likely to receive the vaccination and to trust health experts(Lucia, Kelekar, and Afonso 2021b). People who are more hesitant about vaccination are less likely to obtain information about COVID-19 from traditional and authoritative sources and have similar levels of mistrust in these sources than those who accepted the vaccine(Murphy et al. 2021b). Information provision is crucial to ensure trust in scientific evidence and to form norms to take collective action to mitigate the pandemic(Allcott 2011; Allcott and Knittel 2019; Sasaki, Kurokawa, and Ohtake 2021; Sasaki, Saito, and Ohtake 2021). Therefore, researchers have studied what kind of message, information, education, and social campaigns regarding vaccination reduce hesitancy(Brita Roy, Vineet Kumar, and Arjun Venkatesh 2020; Feleszko et al. 2021). To reduce hesitancy, it may be effective to provide information about the subjective evaluation of the effectiveness of the vaccine, subjective well-being, and mental health.

It is worth analyzing the influence of vaccination on vaccinated people's perceptions of COVID-19, subjective well-being, and mental health. Furthermore, the impact of unexpected shocks, such as COVID-19, differ between males and females(Mohapatra 2021; Yamamura and Tsustsui 2021a, 2021b; Yerkes et al. 2020). To illustrate, the Japanese government's calling for preventive behaviors is less effective for men(Muto et al. 2020). Therefore, males are less likely to change their lifestyles (Yamamura and Tsustsui 2021b). However, compared with males, females were less likely to be hesitant to receive the COVID-19 vaccine(Feleszko et al. 2021; Murphy et al. 2021b). That is, women are more sensitive to COVID-19. These results are consistent with the argument that males are more likely to be overconfident than females(Barber and Odean 2001). Hence, examining gender differences in the effect of vaccination on perceptions and mental health is valuable.

To this end, using monthly individual-level panel data, we investigated how vaccinated people change their perceptions of COVID-19, subjective well-being, and metal health in Japan.

The major findings were as follows: (1) vaccinated people perceived a lower probability of infection than before vaccination. (2) Vaccinated females improved their subjective well-being and mental health, whereas vaccinated males did not change their subjective well-being and mental health. Therefore, providing information about the effect of vaccination on female mental health improvement may increase their motivation to be vaccinated.

# 2. Materials and methods

## 2.1. Data collection

The research company INTAGE, which has sufficient experience in academic research, was commissioned to conduct an internet survey for this study. Individuals registered with INTAGE were recruited as the participants in our project. The sampling method was designed to collect a representative sample of the Japanese population in terms of gender, age, and residential area. However, we restricted Japanese citizens aged–16-79 for the survey because other people were difficult to recruit.

INTAGE conducted internet surveys repeatedly for 15 separate times ("waves") almost every month with the same individuals to construct the panel data. However, in the exceptional period between July 2020 and September 2020, the surveys could not be conducted because of a shortage of research funds. The surveys were resumed after receiving additional funds in October 2020.

The first wave of queries was conducted in the early stage of COVID-19 from March 13 to March 16, 2020. We aimed to collect around 4,000 respondents, distributed to 7,965, and collected 4,359 observations with a response rate of 54.7 %. Respondents from the first wave were targeted in subsequent waves to record how the same respondent changed their perceptions and behaviors during the COVID-19 pandemic. During the study period, until the 15[th] wave was conducted on August 27, 2021, although there were some attritions, the response rate exceeded 83 % at any wave. Accordingly, the total number of observations used in this study was 54,007. In this study, we report results based on unbalanced panel data.

## 2.2.    Ethical considerations

Our study was performed according to relevant guidelines and regulations. The ethics committee of Osaka University approved all survey procedures, and informed consent was obtained from all participants.

All survey participants provided their consent to participate in the anonymous online survey. After being informed about the purpose of the study and their right to quit the survey, participants agreed to participate. The completion of the entire questionnaire was considered to indicate the participants' consent.

## 2.3.    Measurements

Table 1 presents a description of the variables and the mean difference test between men and women. The survey questionnaire included basic questions about demographics, such as age, gender, and educational background. As the main variables, the respondents were asked questions concerning perceptions about COVID-10 as follows:

*According to you, what is the probability (%) of you getting infected with SARS-COV-2 within a month from now? "What percentage do you think the probability of your taking the COVID-19? Choose a percentage from 0 to 100 (%)"*

*"How serious are your symptoms if you are infected with the novel coronavirus?*

*Choose from six choices: 1 (very small influence) to 6 (death)."*

The answers to the questions served as proxies for the subjective probability of contracting COVID-19 and their perceptions of the severity of COVID-19. Larger values indicated that respondents are more likely to perceive a higher risk of COVID-19. Further, as key variables to reflect subjective well-being and mental health, we also asked the following questions:

Concerning subjective wellbeing:

*"How happy do you feel now?*

*Please answer on a scale from 1 (very unhappy) to 11 (very happy)."*

Concerning mental health:

*In the last two weeks, to what extent have you felt anger, fear, and anxiety? Please indicate from 1 (I have not felt the emotion the slightest bit) to 5 (I have felt the emotion more strongly than ever).*

Larger values indicated that the respondents' mental health was worse. Apart from subjective values, an important question was to ask the respondents whether they took the first shot of the vaccine against COVID-19 and whether they had completed the second shot after the 12th wave. Using the data of the second shot, we also defined dummy variables, VACCINE SECOND_1 to VACCINE SECOND_4, representing the time when they were vaccinated.

Table 1 suggests that the mean values of *PROB_COVID19* and *SEVER_COVID19* for females were significantly larger than for males. Hence, females are more likely to perceive COVID-19 as risky than males. Mean values of *HAPPY, FEAR, ANXIETY,* and *ANGER* were significantly larger for women than for men. This implies that females were happier than males, even during the COVID-19 pandemic, whereas females' metal health was worse than for males. The low level of female mental health is consistent with the observation that women's suicide rates increased after the spread of COVID-19(Sakamoto et al. 2021). In contrast to subjective values, there were differences in *vaccine first* and *vaccine second* between men and women.

**Table 1. Definitions of key variables and its mean difference test between males and females.**

| Variables | Definition | Male (1) | Female (2) | (2) – (1) |
|---|---|---|---|---|
| *PROB_COVID19* | What percentage do you think the probability of your taking the COVID-19? 0 to 100 (%) | 18.9 | 22.0 | 3.11*** |
| *SEVER_COVID19* | How serious are your symptoms if you are infected with the novel coronavirus? Choose from 6 choices. 1 (very small influence) to 6 (death) | 3.53 | 3.62 | 0.09*** |
| *HAPPY* | To what degree are you currently feeling happiness? Please answer on a scale from 1 (very unhappy) to 11 (very happy) | 6.59 | 7.04 | 0.45*** |
| *FEAR* | How much have you felt the emotions of fear? Please answer in a scale from 1 (I have not felt this emotion at all) to 5 (I have felt this emotion strongly). | 2.94 | 3.20 | 0.25*** |
| *ANXIETY* | How much have you felt the emotions of anxiety? Please answer in a scale from 1 (I have not felt this emotion at all) to 5 (I have felt this emotion strongly). | 3.15 | 3.43 | 0.29*** |
| *ANGER* | How much have you felt the emotions of anger? Please answer in a scale from 1 (I have not felt this emotion at all) to 5 (I have felt this emotion strongly). | 2.94 | 3.03 | 0.09*** |
| *VACCINE FIRST* | Did you take the first shot (but not yet the second one)? 1 (Yes) or 0 (No) | 0.38 | 0.38 | 0.002 |
| *VACCINE SECOND* | Did you take the second shot? 1 (Yes) or 0 (No) | 0.06 | 0.06 | 0.001 |
| *VACCINE SECOND_1* | 1 if they took the second shot this month, 0 otherwise | 0.03 | 0.03 | 0.0002 |
| *VACCINE SECOND_2* | 1 if they took the second shot the last month, 0 otherwise | 0.02 | 0.02 | 0.001 |
| *VACCINE SECOND_3* | 1 if they took the second shot two months ago, 0 otherwise | 0.006 | 0.006 | −0.0003 |
| *VACCINE SECOND_4* | 1 if they took the second shot three months ago, 0 otherwise | 0.001 | 0.001 | 0.00003 |

Note: ***p<0.01

      **p<0.05

      *p<0.10

The Japanese government began vaccination in February 2021(Japan Times 2021b). During the early period of vaccination, the initial group receiving the shot was strictly restricted to health workers. Vaccination for general older people aged 65 and over has been implemented since April 2021. Accordingly, 75 % of older people were vaccinated in July 2021(Japan Times 2021a). Subsequently, COVID-19 vaccination programs began at workplaces and campuses where workers and students received vaccinations in June(Japan Times 2021c).

In our project, we started asking about vaccines against Sars-CoV-2 at the 7[th] wave

(December 2020) when vaccination, globally, first started in Israel. The question of which data were used in this study, asking whether respondents received their first and second shot, appeared from the 12[th] wave conducted in May 2021. At that time, the completed ratios of the first and second shots were 5.24 % and 0.59 % of the total nation, respectively. We created the dummy variables, *VACCINE SECOND_1* to *VACCINE SECOND_4,* to capture the timing of the second shot. Although their mean values may seem quite low, this is because we set the value of these variables before the 12[th] wave at zero, reflecting the reality in Japan. On April 23, 2021, as of the 11[th] wave, the inoculation rates were 0.23 % and 0.00 % for the first and second shots, respectively.

To determine the change in the vaccination rate, Table 2 shows the percentages of vaccinated people in the whole sample, male sample, and female sample in each wave. Table 1 reports the aggregated values containing both the first and second shot vaccinated people, regardless of vaccination time point. Inevitably, the percentage of vaccinated individuals is expected to increase over time. In line with this inference, Table 2 shows that the percentage of vaccinated people rapidly increased from 8.2 % in May 2021 to 64.2 % in September in our sample. This rate is almost the same as that of 65.2 % in September in a country-wide sample(Department of Medical Genome Sciences 2021). Thus, the data of this study reflect the actual situation in Japan. Further, a similar tendency was observed when we used a sub-sample of males and females.

**Table 2. Percentage of those who took the COVID-19 vaccine.**

| Waves | Dates | All % | Males % | Females % | First shot % | Second shot % |
|---|---|---|---|---|---|---|
| 1 | March 13–16, 2020 | 0 | 0 | 0 | | |
| 2 | March 27–30, 2020 | 0 | 0 | 0 | | |
| 3 | Apr. 10–13, 2020 | 0 | 0 | 0 | | |
| 4 | May 8–11, 2020 | 0 | 0 | 0 | | |
| 5 | June 12–15, 2020 | 0 | 0 | 0 | | |
| 6 | Oct 23–28, 2020 | 0 | 0 | 0 | | |
| 7 | Dec 4–8, 2020 | 0 | 0 | 0 | | |
| 8 | Jan. 15–19, 2021 | 0 | 0 | 0 | | |
| 9 | Feb. 17–22, 2021 | 0 | 0 | 0 | | |
| 10 | Mar. 24–29, 2021 | 0 | 0 | 0 | | |
| 11 | Apr. 23–26, 2021 | 0 | 0 | 0 | | |
| 12 | May 28–31, 2021 | 8.2 | 8.2 | 8.2 | 5.24 | 0.63 |
| 13 | June 25–30, 2021 | 25.1 | 24.3 | 25.9 | 19.63 | 8.01 |
| 14 | July 30–Aug 4, 2021 | 50.0 | 48.5 | 51.4 | 39.48 | 26.86 |
| 15 | Aug 27–Sep. 1, 2021 | 64.2 | 63.7 | 64.7 | 51.29 | 91.23 |

Note: We did not distinguish respondents who took only the first shot from those who took the second shot.

Fig 1 illustrates perceptions about COVID such as *PROB_COVID19* and *SEVER_COVID19* from the 1st to the 15th waves for vaccinated and non-vaccinated groups. In the Figure, the vaccinated group is defined as those who were vaccinated at any time point during our observation period. Further, the group includes both people who received the second shot and those who only received the first shot. For example, an individual who had their first shot in the 15th wave was included in the vaccinated group.

Nobody was vaccinated before the 12th wave, as shown in the left part of the vertical line in Fig 1. Fig 1 suggests how people who were not vaccinated behaved differently from vaccinated people in the period when the vaccine was not distributed.

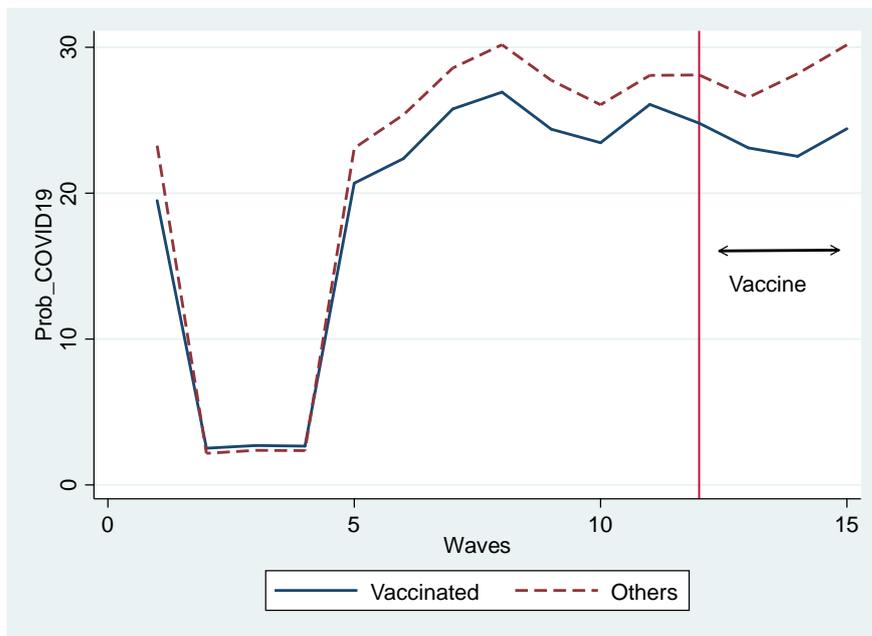

(a) Prob_COVID-19

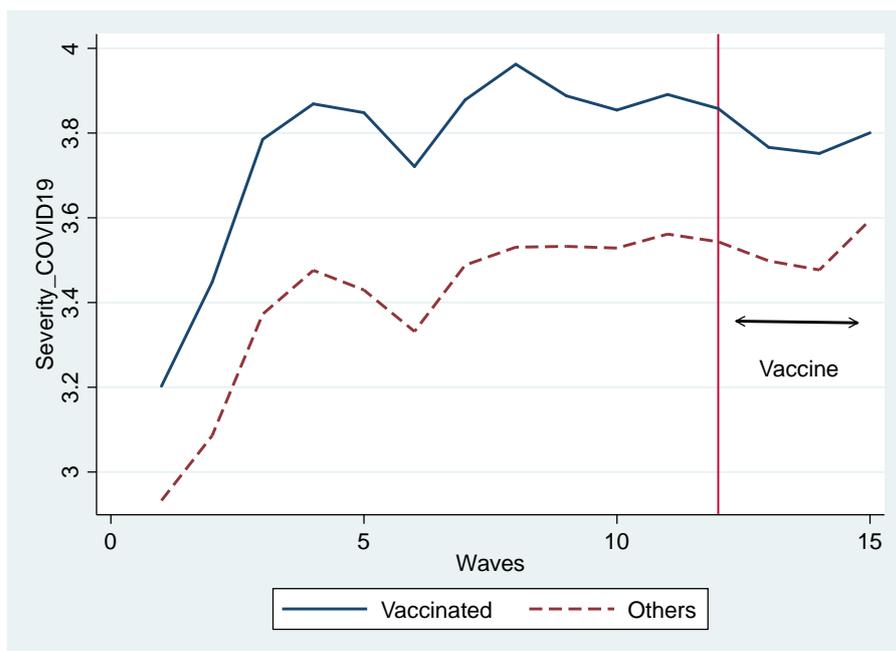

(b) Severity of COVID-19

Fig 1.

  Fig 1(a) indicates that the vaccinated group perceived the probability of getting COVID-19 to be lower than that of the non-vaccinated group, even before distribution of the vaccine. The trends of both groups were similar. During the first declaration of a state of emergency in all parts of Japan from the 3rd to 4th waves (April 7–May 27, 2020), the perceived probability drastically declined and remained at the lowest level. After the first declaration was terminated, its level increased to a level higher than that before the declaration. Later, its level did not remarkably change even though a state of emergency was declared and called off repeatedly four times. However, it should be noted that the gap between the groups increased after 2021 (the 8th wave). Contrastingly, Fig 1. (b) indicates that the subjective severity of COVID-19 was consistently higher in the vaccinated than in the non-vaccinated group. Even during the first declaration of a state of emergency, subjective severity increased drastically. After termination, the level of subjective severity was relatively stable. After distribution of the vaccine, the gap between the groups was reduced. The only similarity between Fig 1 (a) and (b) is that the levels of both variables increase in the non-vaccinated group.

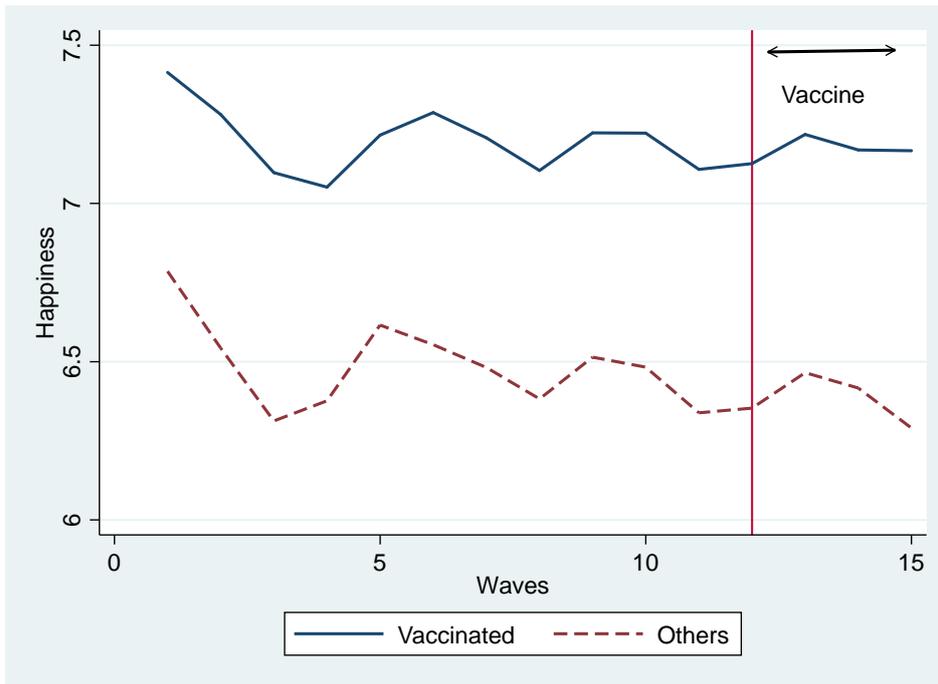

(a) Happiness

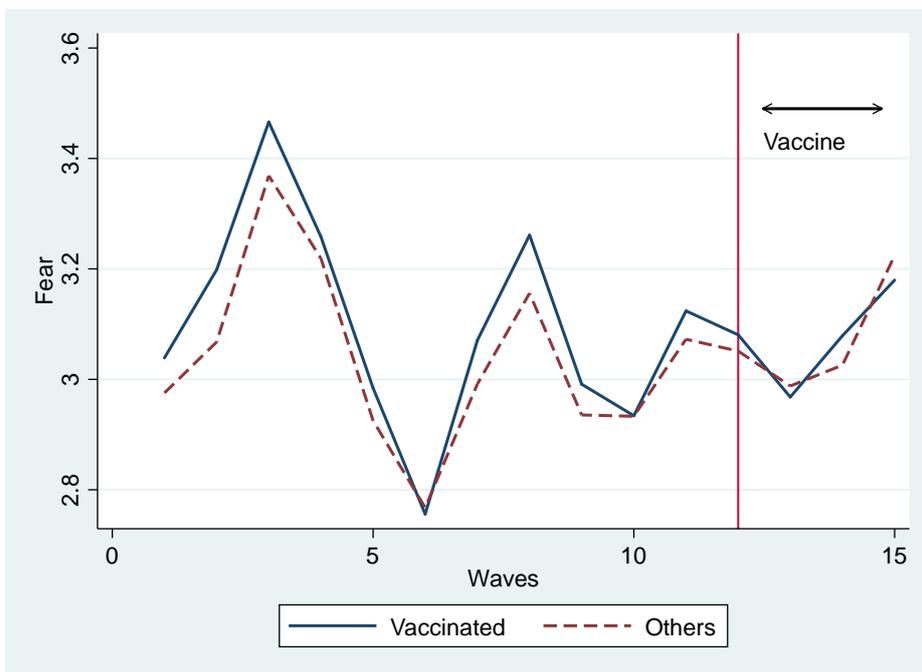

(b) Fear

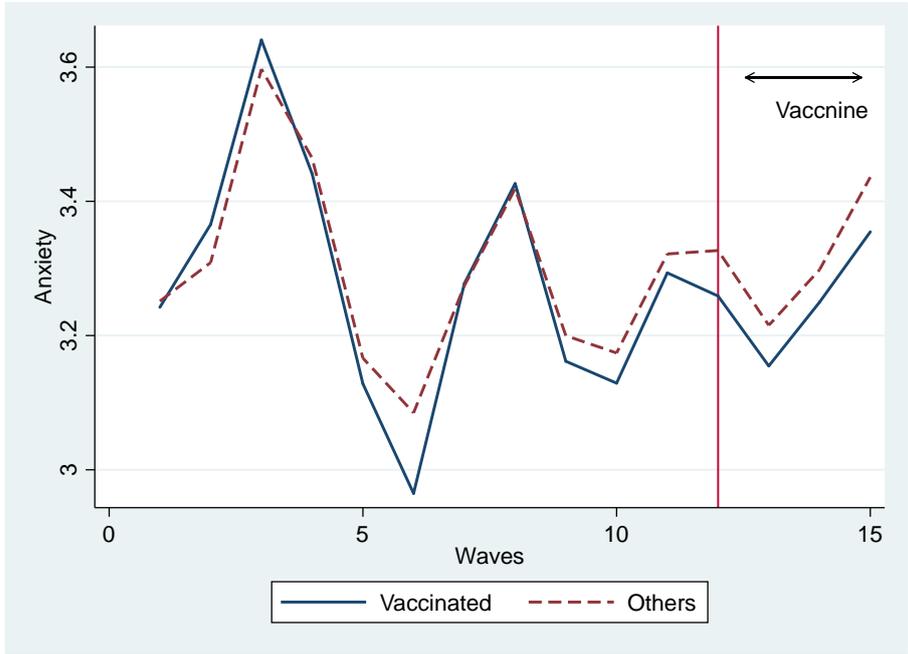

(c) Anxiety

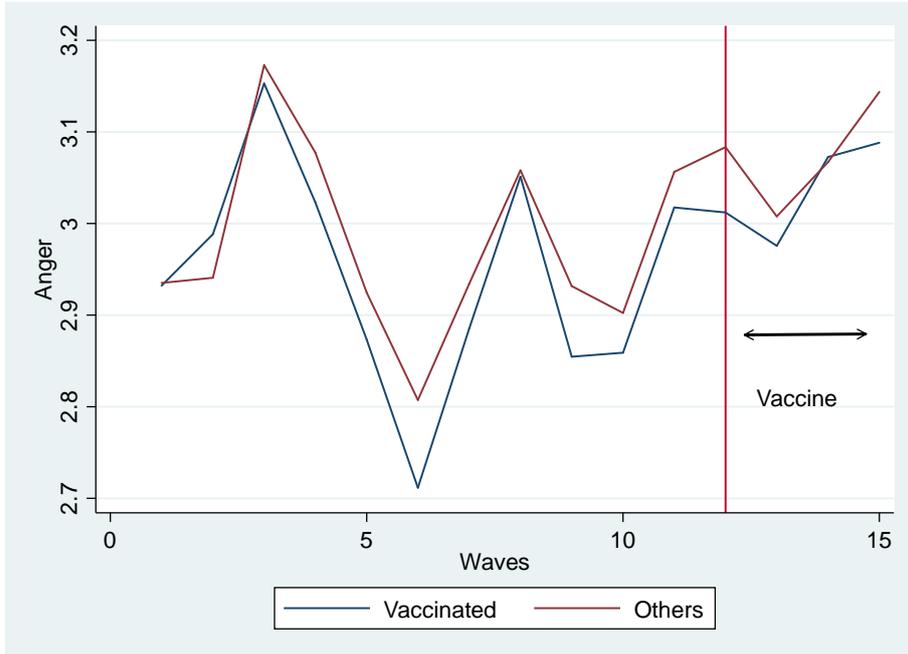

(d) Anger

Fig. 2.

In the panels in Fig. 2, we illustrate *HAPPY, FEAR, ANXIETY,* and *ANGER* in the same manner as in Fig.1. The *HAPPY* level was consistently higher in the vaccinated than in the non-vaccinated group. During the first state of the emergency period, the level of *HAPPY* declined, whereas it increased after its termination. There were cyclical changes throughout the period, although the amplitude of the changes was not large compared with the gap between the groups. After the initiation of the vaccine, the happiness of the non-vaccinated people declined, whereas that of vaccinated people was stable. Hence, the gap in happiness levels between the groups widened.

Regarding the three proxies for mental health, *FEAR, ANXIETY,* and *ANGER* changed similarly. These negative feelings increased drastically after entering the first state of emergency and dropped to the lowest level. Similar to *HAPPY*, the three emotions showed cyclical changes. However, the gap between these three proxies did not show a remarkable systematic change.

Overall, in Figs. 1 and 2 (a), vaccination leads people to have more positive views. Meanwhile, in Figs. 2 (b), 2 (c), and 2 (d), we did not observe an apparent effect of vaccination before and after the distribution of the COVID-19 vaccine. However, these observations reflect changes in the mean values when various factors are not controlled. For a closer examination of the effects of vaccination, we examined the fixed-effects regression model.

## 2.4. Methods

A fixed effects (FE) regression model was used to control the time-invariant individual characteristics. The estimated function took the following form:

$Y_{it} = \alpha_1 \ VACCINE \ FIRST_{it} + \alpha_2 \ VACCINE \ SECOND\_1_{it} + \alpha_3 \ VACCINE \ SECOND\_2_{it} + \alpha_4 \ VACCINE \ SECOND\_3_{it} + \alpha_5 \ VACCINE \ SECOND\_4_{it} + \alpha_6 \ EMERGENT_{it} + k_t + m_i + u_{it},$

In this formula, $Y_{it}$ represents the dependent variable for individual i and wave t. The dependent variable is a proxy for the perception of COVID-19, *PROB_COVID19*, and *SEVER_COVID19*. To examine mental health, *HAPPY, FEAR, ANXIETY*, and *ANGER* were used as the dependent variables. The regression parameters are denoted as α. The error term is denoted by *u*.

$k_t$ represents the effects of different time points. This is controlled by including 14 wave dummy variables, where the 1st wave is taken as the reference group. Fourteen wave dummy variables captured various shocks that occurred simultaneously throughout Japan at each time point. For instance, the state of emergency was declared four times during the study period. The time-invariant individual-level fixed effects are represented by $m_i$. The FE model controls various individual characteristics that do not change over time. Therefore, the model controls for various time-invariant factors including sex, birth year, and experiences in past personal history, such as educational background.

Key independent variables were dummy variables for vaccination; *VACCINE FIRST* represents the effect of the first shot. The Japanese government approved only the Pfizer-BioNTech and the Moderna vaccines. The first vaccinated persons were obliged to take the second shot within a month to ensure the vaccine's effectiveness. This rule was applied to Pfizer, BioNTech, and Moderna vaccines. That is, the vaccine was effective enough for those who took only the first shot. Hence, we should scrutinize the effect of vaccination by considering the first and second shots separately. Further, it is noteworthy to investigate the effect of the vaccine on perceptions and mental health changes over time. To this end, we incorporated four dummy variables: *VACCINE SECOND_1, VACCINE SECOND_2, VACCINE SECOND_3,* and *VACCINE SECOND_4.*

Vaccination is expected to reduce the probability of contracting COVID-19 and its severity. Hence, the expected sign of the dummy variables for vaccination was negative for these variables. Moreover, vaccination is anticipated to improve negative emotions. Therefore, the coefficients of *FEAR, ANXIETY*, and *ANGER* were expected to show a negative sign, whereas *HAPPY* was anticipated to exhibit a positive sign.

Concerning control variables, in Japan, declarations of a state of emergency significantly influenced individuals' behaviors(Yamamura and Tsustsui 2021c; Yamamura and Tsutsui 2020). The timing of the declarations varied according to the area where one resided. Therefore, the effect of the declaration could not be captured by wave dummy variables. Accordingly, we included *EMERGENT* to control for this effect. We also controlled for the following factors: the number of persons infected with COVID-19 and deaths caused by COVID-19 in residential areas at each time point, although their results were not reported due to space limits.

In addition to estimation using the whole sample, we report the estimates by dividing the sample according to the respondent's sex to compare the effect of vaccination between males and females.

# 3. Results

## 3.1. Full sample estimations

Tables 3 and 4 report the estimation results of the FE model using the entire sample. We begin by interpreting the key vaccination dummy variables to capture the effect of the first vaccination, *VACCINE_FRIST,* and to capture the effect of the second vaccination and its duration effect, *VACCINE_SECOND_1, VACCINE_SECOND_2, VACCINE_SECOND_3,* and *VACCINE_SECOND_4*. Concerning the perceptions of COVID-19, in the estimation of *PROB_COVID19,* all vaccine dummy variables showed the expected negative sign with statistical significance at the 1 % level. The absolute values of their coefficients were 1.25 for *VACCINE_FRIST*. This implies that respondents perceived that the probability of getting an infection was reduced by 1.25 % directly after they got the first shot. The values for the second vaccination dummy variables were 4.37, 5.08, 4.98, and 4.82 for *VACCINE_SECOND_1, VACCINE_SECOND_2, VACCINE_SECOND_3,* and *VACCINE_SECOND_4,* respectively. In our interpretation, an individual's perceived probability was lower by 4.37 % directly after the second shot than before the first shot. The effect of the second shot increased to 5.08 % in the next month, but slightly decreased to 4.98 % after two months, and then to 4.82 % after three months. Overall, the effect of the second shot was about four times larger than that of the first shot and persisted over time. Estimations for *SEVER_COVID19* also showed similar results, although *VACCINE_SECOND_4* showed neither the negative sign nor statistical significance. This means that an individual's perception of the severity of COVID-19 returned to the level before vaccination after three months since they took the second shot. The absolute value of the coefficient of *VACCINE_FRIST* was 0.04, meaning that the perceived severity of COVID-19 decreased by 0.04 points on a 5-point scale directly after they received the first shot. The values were 0.173, 0.181, and 0.142 for *VACCINE_SECOND_1, VACCINE_SECOND_2,* and *VACCINE_SECOND_3,* respectively. This can be interpreted as the perceived severity of COVID-19 decreasing by around 0.14-0.18 points on a 5-point scale after they got the second shot compared to before they were vaccinated. Similar to the results of *PROB_COVID19*, the degree of the second shot effect was approximately four times larger than that of the first shot. These observations reasonably reflect that the second shot substantially leads to the vaccine being more effective. The effect was at the peak one month after the second shot, which is similar to the results of PROB_COVID19.

As for subjective happiness, in column (3) of *HAPPY*, we observed the expected positive sign for vaccination dummy variables with the exception of *VACCINE_SECOND_4*. However, statistical significance was observed only for *VACCINE_SECOND_1* and *VACCINE_SECOND_3*. Therefore, the positive effect of vaccination was observed to a certain extent but was not robust. Concerning mental health estimations, for estimations of *FEAR,* all vaccine dummy variables exhibited the expected negative sign and were statistically significant. The absolute value of the coefficient of *VACCINE_FRIST* was 0.04, implying that the level of fear decreased by 0.04 points on a 5-point scale directly after they received the first shot. The values were 0.08, 0.09, and 0.13 for *VACCINE_SECOND_1, VACCINE_SECOND_2,* and *VACCINE_SECOND_3,* respectively. Therefore, the second shot effect was approximately two to three times larger than that of the first shot. However, the degree of increase from the first to the second shot was smaller than the estimations for *PROB_COVID19* and *SEVER_COVID19*. The estimation results for *ANXIETY* were similar to XXX, although *VACCINE_FIRST* and *VACCINE_SECOND_4* were not significant. In contrast, in the estimation of *ANGER*, all vaccination dummies were not statistically significant, implying that the vaccination weakened fear but not anger.

**Table 3. FE model. Dependent variables are perceptions of COVID-19 and mental health. Sample including males and females.**

| | (1) PROB_ COVID19 | (2) SEVER_ COVID19 | (3) HAPPY | (4) FEAR | (5) ANXIETY | (6) ANGER |
|---|---|---|---|---|---|---|
| *VACCINE FIRST* | −1.248*** | −0.044** | 0.024 | −0.041** | −0.026 | −0.011 |
| | (0.44) | (0.02) | (1.07) | (0.02) | (0.02) | (0.02) |
| *VACCINE SECOND_1* | −4.369*** | −0.173*** | 0.063** | −0.078*** | −0.059*** | −0.023 |
| | (0.49) | (0.02) | (0.03) | (0.02) | (0.02) | (0.02) |
| *VACCINE SECOND_2* | −5.084*** | −0.181*** | 0.042 | −0.092*** | −0.064** | 0.029 |
| | (0.59) | (0.03) | (0.05) | (0.03) | (0.03) | (0.03) |
| *VACCINE SECOND_3* | −4.980*** | −0.142* | 0.160** | −0.129*** | −0.139*** | −0.008 |
| | (0.73) | (0.07) | (0.07) | (0.04) | (0.04) | (0.05) |
| *VACCINE SECOND_4* | −4.821*** | 0.041 | −0.188 | −0.017* | −0.073 | −0.124 |
| | (1.71) | (0.08) | (0.15) | (0.09) | (0.07) | (0.09) |
| *EMERGENT* | −0.007 | −0.005 | −0.015 | 0.039** | 0.039*** | 0.013** |
| | (0.25) | (0.01) | (0.02) | (0.02) | (0.01) | (0.01) |
| *WAVE 1* | | | <Default> | | | |
| *WAVE 2* | −19.385*** | 0.184*** | −0.173*** | 0.119*** | 0.083*** | 0.023 |
| | (0.48) | (0.02) | (0.02) | (0.02) | (0.02) | (0.02) |
| *WAVE 3* | −19.033*** | 0.494*** | −0.383*** | 0.390*** | 0.349*** | 0.215*** |
| | (0.53) | (0.02) | (0.03) | (0.02) | (0.02) | (0.02) |
| *WAVE 4* | −19.107*** | 0.582*** | −0.349*** | 0.181*** | 0.155*** | 0.081*** |
| | (0.55) | (0.05) | (0.03) | (0.03) | (0.02) | (0.03) |
| *WAVE 5* | 0.497 | 0.532*** | −0.149*** | −0.061** | −0.106*** | −0.036 |
| | (0.39) | (0.01) | (0.03) | (0.02) | (0.02) | (0.03) |
| *WAVE 6* | 2.622*** | 0.416*** | −0.140*** | −0.253*** | −0.227*** | −0.179*** |
| | (0.49) | (0.02) | (0.03) | (0.01) | (0.02) | (0.02) |
| *WAVE 7* | 5.726*** | 0.563*** | −0.216*** | −0.011 | 0.017 | −0.034 |
|  | (0.53) | (0.02) | (0.03) | (0.02) | (0.02) | (0.02) |
| *WAVE 8* | 6.957*** | 0.627*** | −0.326*** | 0.175*** | 0.146*** | 0.094*** |
|  | (0.53) | (0.02) | (0.04) | (0.03) | (0.03) | (0.02) |
| *WAVE 9* | 4.486*** | 0.597*** | −0.180*** | −0.076*** | −0.101*** | −0.066*** |
|  | (0.52) | (0.02) | (0.03) | (0.03) | (0.03) | (0.02) |
| *WAVE 10* | 3.474*** | 0.588*** | −0.194*** | −0.084*** | −0.109*** | −0.061** |
|  | (0.46) | (0.02) | (0.04) | (0.02) | (0.02) | (0.02) |
| *WAVE 11* | 5.647*** | 0.609*** | −0.308*** | 0.070*** | 0.036* | 0.081*** |
|  | (0.45) | (0.02) | (0.03) | (0.02) | (0.02) | (0.02) |
| *WAVE 12* | 5.074*** | 0.588*** | −0.294*** | 0.035 | 0.018*** | 0.097*** |
|  | (0.40) | (0.02) | (0.03) | (0.02) | (0.02) | (0.02) |
| *WAVE 13* | 4.125*** | 0.551*** | −0.231*** | −0.018 | −0.057*** | 0.058*** |
|  | (0.38) | (0.02) | (0.03) | (0.02) | (0.02) | (0.02) |
| *WAVE 14* | 5.624*** | 0.579*** | −0.298*** | 0.069** | 0.035*** | 0.121*** |
|  | (0.55) | (0.02) | (0.03) | (0.03) | (0.03) | (0.03) |
| *WAVE 15* | 7.934*** | 0.674*** | −0.343*** | 0.221*** | 0.150*** | 0.143*** |
|  | (0.60) | (0.03) | (0.05) | (0.04) | (0.03) | (0.02) |
| Adj $R^2$ | 0.57 | 0.67 | 0.76 | 0.56 | 0.57 | 0.50 |
| Obs. | 54,007 | 54,007 | 54,007 | 54,007 | 54,007 | 54,007 |

**Note**: Numbers within parentheses are robust standard errors clustered in the residential prefectures. The model includes the number of deaths and infected persons in residential prefectures at the time of the surveys. However, these results have not been reported. These are included, although the results have not been reported.

***p<0.01

**p<0.05

*p<0.10

Table 4 shows alternative specifications where a second shot dummy variable was used to examine the effect of the second shot vaccination instead of using four dummy variables to capture the timing of the second shot. In Table 4, we focus on whether respondents completed the second shot. Hence, Table 4 reports the key variables, although the set of control variables are the same as in Table 3. Results were similar as shown in Table 3. The significant expected sign of *VACCINE SECOND* was observed in columns (1)-(5), but no statistical significance was observed in column (6). Except for column (6), the absolute values of coefficient and statistical significance were larger for *VACCINE*

*SECOND* than *VACCINE FIRST*. Therefore, individuals have more optimistic views about COVID-19 and their subjective well-being and mental health improved after they took the second shot.

**Table 4. FE model. Dependent variables are perceptions of COVID-19 and mental health. Sample including males and females. (Alternative specification)**

|  | (1) *PROB_ COVID19* | (2) *SEVER_ COVID19* | (3) *HAPPY* | (4) *FEAR* | (5) *ANXIETY* | (6) *ANGER* |
|---|---|---|---|---|---|---|
| *VACCINE FIRST* | −1.253*** (0.43) | −0.044** (0.02) | 0.023 (0.02) | −0.040** (0.02) | −0.025 (0.02) | −0.009 (0.02) |
| *VACCINE SECOND_1* | −4.676*** (0.45) | −0.169*** (0.02) | 0.058** (0.03) | −0.085*** (0.02) | −0.064*** (0.02) | −0.004 (0.02) |
| Adj R$^2$ | 0.57 | 0.67 | 0.76 | 0.56 | 0.57 | 0.50 |
| Obs. | 54,007 | 54,007 | 54,007 | 54,007 | 54,007 | 54,007 |

**Note**: Numbers within parentheses are robust standard errors clustered in the residential prefectures. The set of control variables used in Table 3 is included, although the results are not reported.

***p<0.01

**p<0.05

*p<0.10

## 3.2. Sub-sample estimations (Male vs Female groups)

Tables 5 and 6 report the results based on a sub-sample of males, while Tables 7 and 8 present the results based on a sub-sample of females. Here, we focused on key variables, although the same set of control variables used in Table 3 was included. The results of the perceptions of COVID-19, *PROB_COVID19,* and *SEVER_COVID19* in Table 5 show similar results to those in Table 3. Vaccination dummy variables showed the expected negative sign in all results and was statistically significant in most of the results. Comparatively, for results of *HAPPY, FEAR,* and *ANGER*, we did not observe statistical significance with the exception of VACCINE SECOND_1 in columns (6). This implies that males were more likely to have an optimistic view about COVID-19, whereas their mental health did not improve by receiving the vaccine. This tendency is consistently observed in Table 6.

**Table 5. FE model: Dependent variables are perceptions of COVID-19 and mental health. Male sample.**

|  | (1) *PROB_COVID19* | (2) *SEVER_COVID19* | (3) *HAPPY* | (4) *FEAR* | (5) *ANXIETY* | (6) *ANGER* |
|---|---|---|---|---|---|---|
| *VACCINE FIRST* | −1.876*** (0.47) | −0.017 (0.02) | 0.005 (0.05) | 0.001 (0.02) | 0.006 (0.03) | 0.009 (0.03) |
| *VACCINE SECOND_1* | −3.684*** (0.60) | −0.163*** (0.03) | 0.019 (0.05) | −0.038 (0.03) | −0.052* (0.03) | −0.016 (0.03) |
| *VACCINE SECOND_2* | −5.018*** (0.81) | −0.193*** (0.05) | −0.043 (0.07) | −0.044 (0.04) | −0.032 (0.05) | 0.040 (0.04) |
| *VACCINE SECOND_3* | −4.890*** (1.06) | −0.174* (0.09) | 0.133 (0.10) | −0.023 (0.06) | −0.063 (0.06) | 0.104 (0.07) |
| *VACCINE SECOND_4* | −4.123** (2.05) | − 0.045 (0.07) | −0.207 (0.16) | −0.059 (0.13) | −0.055 (0.12) | −0.143 (0.13) |
| Adj $R^2$ | 0.57 | 0.65 | 0.77 | 0.56 | 0.57 | 0.57 |
| Obs. | 27,316 | 27,316 | 27,316 | 27,316 | 27,316 | 27,316 |

Note: Numbers within parentheses are robust standard errors clustered in the residential prefectures. The set of control variables used in Table 3 is included, although the results are not reported.

***p<0.01

**p<0.05

*p<0.10

**Table 6. FE model: Dependent variables are perceptions of COVID-19 and mental health. Male sample. (Alternative specification)**

|  | (1) PROB_COVID19 | (2) SEVER_COVID19 | (3) HAPPY | (4) FEAR | (5) ANXIETY | (6) ANGER |
|---|---|---|---|---|---|---|
| *VACCINE FIRST* | −1.876*** (0.47) | −0.015 (0.02) | 0.0003 (0.04) | 0.003 (0.02) | 0.008 (0.03) | 0.010 (0.03) |
| *VACCINE SECOND 1* | −4.23*** (0.59) | −0.167*** (0.04) | −0.005 (0.05) | −0.034 (0.03) | −0.042 (0.03) | 0.014 (0.03) |
| Adj R² | 0.57 | 0.65 | 0.77 | 0.56 | 0.57 | 0.57 |
| Obs. | 27,316 | 27,316 | 27,316 | 27,316 | 27,316 | 27,316 |

**Note**: Numbers within parentheses are robust standard errors clustered in the residential prefectures. The set of control variables used in Table 3 is included, although the results are not reported.

***p<0.01

**p<0.05

*p<0.10

Table 7 indicates that the coefficients of vaccination dummy variables are negative and statistically significant in most cases in the estimations of *PROB_COVID19* and *SEVER_COVID19*. Further, except for column (6) where *ANGER* results are shown, most of the vaccination results showed the expected sign and statistical significance. We also observed consistent results as shown in Table 8.

Tables 5–8 jointly reveals the gender differences in the vaccination effect on subjective well-being and mental health. Thus, vaccination had a positive influence on women's but not on males' mental health.

**Table 7. FE model: Dependent variables are perceptions of COVID-19 and mental health. Female sample.**

|  | (1) | (2) | (3) | (4) | (5) | (6) |
|---|---|---|---|---|---|---|

|  | PROB_ COVID19 | SEVER_ COVID19 | HAPPY | FEAR | ANXIETY | ANGER |
|---|---|---|---|---|---|---|
| VACCINE FIRST | −0.617 (0.69) | −0.072** (0.03) | 0.047 (0.04) | −0.084*** (0.03) | −0.060** (0.03) | −0.030 (0.02) |
| VACCINE SECOND_1 | −5.068*** (0.64) | −0.184*** (0.03) | 0.112*** (0.03) | −0.116*** (0.02) | −0.066** (0.03) | −0.029 (0.03) |
| VACCINE SECOND_2 | −5.245*** (0.94) | −0.171*** (0.04) | 0.131** (0.06) | −0.139*** (0.04) | −0.097*** (0.03) | 0.017 (0.04) |
| VACCINE SECOND_3 | −5.043*** (1.33) | −0.109 (0.08) | 0.189* (0.01) | −0.241*** (0.05) | −0.217*** (0.05) | −0.127* (0.07) |
| VACCINE SECOND_4 | −5.521** (2.47) | 0.125 (0.11) | −0.157 (0.21) | −0.280 (0.18) | −0.098 (0.09) | −0.106 (0.14) |
| Adj R$^2$ | 0.56 | 0.68 | 0.73 | 0.54 | 0.55 | 0.48 |
| Obs. | 26,691 | 26,691 | 26,691 | 26,691 | 26,691 | 26,691 |

Note: Numbers within parentheses are robust standard errors clustered in the residential prefectures.

The set of control variables used in Table 3 is included, although the results are not reported.

***p<0.01

**p<0.05

*p<0.10

**Table 8. FE model: Dependent variables are perceptions of COVID-19 and mental health. Female sample. (Alternative specification)**

|  | (1) PROB_ COVID19 | (2) SEVER_ COVID19 | (3) HAPPY | (4) FEAR | (5) ANXIETY | (6) ANGER |
|---|---|---|---|---|---|---|
| VACCINE FIRST | −0.624 (0.69) | −0.074** (0.03) | 0.049 (0.04) | −0.084*** (0.03) | −0.059* (0.03) | −0.029 (0.02) |
| VACCINE SECOND_1 | −5.151*** (0.63) | −0.174*** (0.03) | 0.127** (0.04) | −0.134*** (0.02) | −0.085*** (0.03) | −0.021 (0.03) |
| Adj R$^2$ | 0.56 | 0.68 | 0.73 | 0.54 | 0.55 | 0.48 |
| Obs. | 26,691 | 26,691 | 26,691 | 26,691 | 26,691 | 26,691 |

**Note**: Numbers within parentheses are robust standard errors clustered in the residential prefectures.

The set of control variables used in Table 3 is included, although the results are not reported.

***p<0.01

**p<0.05

*p<0.10

# 4. Discussion

People's hesitancy for COVID-19 vaccination has hampered the establishment of herd immunity and the termination of the COVID-19 pandemic. People who hesitate to vaccinate are less inclined to access information about COVID-19 from formal and authoritative sources but tend to distrust them (Murphy et al. 2021b). Therefore, it is important to provide effective information that is more acceptable to them and, thus, motivate them to be vaccinated. This holds especially for females because they are more likely to be hesitant to be vaccinated(Feleszko et al. 2021; Murphy et al. 2021b). Providing positive evaluations of vaccination from vaccinated females plays a key role because females with hesitation pay more attention to information from the same sex.

The estimation results made it evident that vaccinated females perceived a lower probability of getting infected and had better subjective well-being and mental health than before vaccination. Providing this information may lead unvaccinated females to more positively view vaccination, which, in turn, will motivate them to be vaccinated. In Japan, female suicide rates have increased during the COVID-19 pandemic. The evidence of this study reveals that vaccination can cure women with mental illness (Mazereel et al. 2021a, 2021b; Siva 2021; Warren et al. 2021). Based on the findings, policymakers should display appropriate messages that are targeted to unvaccinated females.